\documentclass{toc}
\usepackage{url}
\usepackage{eprint}
\usepackage{latexsym}
\usepackage{amsfonts}

\tocdetails{%
   volume=2, number=1, year=2006, firstpage=1,
   received={August 4, 2005},
   published={January XX, 2006}
}

\runningauthor{Robert {\v S}palek, Mario Szegedy}
\runningtitle{All Quantum Adversary Methods are Equivalent}
\copyrightauthor{Robert {\v S}palek, Mario Szegedy}

\long\def\rem#1{}
\def\01{\{0,1\}}
\renewcommand{\O}[1]{\mathrm{O}\mathchoice{\!}{\!}{}{}\left(#1\right)}
\newcommand{\Om}[1]{\Omega\mathchoice{\!}{\!}{}{}\left(#1\right)}
\newcommand{\Th}[1]{\Theta\mathchoice{\!}{\!}{}{}\left(#1\right)}
\newcommand{\eps}{\varepsilon}
\newcommand{\ket}[1]{|#1\rangle}
\newcommand{\Tr}{\mathinner{\mathrm{tr}}}
\newcommand{\diag}{\mathinner{\mathrm{diag}}}
\newcommand{\Cert}[1]{\mathcal{C}\!_f(#1)}
\newcommand{\rownorm}{r}
\newcommand{\colnorm}{c}
\newcommand{\SA}{\mathrm{SA}(f)}
\newcommand{\WA}{\mathrm{WA}(f)}
\newcommand{\SWA}{\mathrm{SWA}(f)}
\newcommand{\KA}{\mathrm{KA}(f)}
\newcommand{\MM}{\mathrm{MM}(f)}
\newcommand{\SMM}{\mathrm{SMM}(f)}
\newcommand{\GSA}{\mathrm{GSA}(f)}
\renewcommand{\mod}{\mathop{\mathrm{mod}}}
\newcommand{\myatop}[2]{\genfrac{}{}{0pt}{}{#1}{#2}}

\newcommand{\appref}[1]{\hyperref[#1]{Appendix~\ref*{#1}}}
\newcommand{\eqnref}[1]{\hyperref[#1]{(\ref*{#1})}}
\newenvironment{shortitemize}
{\begin{itemize}\setlength{\itemsep}{0in}}
{\end{itemize}}

\begin{document}

\begin{frontmatter}[classification=text]
\title{All Quantum Adversary Methods are Equivalent}
\tocpdftitle{All Quantum Adversary Methods are Equivalent}
\tocpdfauthor{Robert Spalek, Mario Szegedy}

\author[spalek]{Robert \v Spalek}
\author[szegedy]{Mario Szegedy}

\begin{abstract}
The quantum adversary method is one of the most versatile lower-bound methods
for quantum algorithms.  We show that all known variants of this
method are equivalent: spectral adversary (Barnum, Saks, and Szegedy, 2003),
weighted adversary (Ambainis, 2003), strong weighted adversary (Zhang, 2005),
and the Kolmogorov complexity adversary (Laplante and Magniez, 2004).  We also present
a few new equivalent formulations of the method.  This shows that there is
essentially \emph{one} quantum adversary method.  From our approach, all known
limitations of these versions of the quantum adversary method easily follow.
\end{abstract}

\tockeywords{Quantum computing, query complexity, adversary lower bounds}
\tocacm{F.1.2, F.1.3}
\tocams{81P68, 68Q17}
\end{frontmatter}

\section{Introduction}

\subsection{Lower-bound methods for quantum query complexity}

In the query complexity model, the input is accessed using oracle queries and
the query complexity of the algorithm is the number of calls to the oracle.
The query complexity model is helpful in obtaining time complexity lower
bounds, and often this is the only way to obtain time bounds in the random
access model.

The first lower-bound method on quantum computation was the hybrid
method of~Bennett, Bernstein, Brassard, and Vazirani~\cite{bbbv:hybrid}
to show an $\Om{\sqrt n}$ lower bound on the quantum database search.  Their
proof is based on the following simple observation: If
the value of function $f$ differs on two inputs $x, y$, then the output
quantum states of any bounded-error algorithm for $f$ on $x$ and $y$ must be
almost orthogonal.  On the other hand, the inner product is $1$ at the
beginning, because the computation starts in a fixed state.  By upper-bounding
the change of the inner product after one query, we lower bound the number of
queries that need to be made.

The second lower-bound method is the polynomial method of~Beals, Buhrman,
Cleve, Mosca, and de Wolf~\cite{bbcmw:polynomialsj}.  It is based on the
observation that the
measurement probabilities can be described by low-degree polynomials in the
input bits.  If $t$ queries have been made, then the degree is at most $2 t$.
Since the measurement probabilities are always inside $[0,1]$, one can apply
degree lower bounds for polynomials to obtain good lower bounds for
quantum query complexity.

The third lower-bound method is the quantum adversary method of
Ambainis~\cite{ambainis:lowerb}.  It extends the hybrid method.
Instead of examining a fixed input pair, Ambainis takes an average
over many pairs of inputs.  In this paper, we study different variants
of the quantum adversary method.

The fourth lower-bound method is the semidefinite programming method of
Barnum, Saks, and Szegedy~\cite{bss:semidef}.  It exactly characterizes
quantum query complexity by a semidefinite program.  The dual of this
program gives a lower bound that encompasses the quantum adversary bound.

\subsection{The variants of the quantum adversary method}

The original version of the quantum adversary method, let us call it
\emph{unweighted},
was invented by Ambainis~\cite{ambainis:lowerb}.  It was successfully
used to obtain the following tight lower bounds: $\Om{\sqrt n}$ for
Grover search~\cite{grover:search}, $\Om{\sqrt n}$ for two-level
And-Or trees (see~\cite{hmw:berror-search} for a matching upper
bound), and $\Om{\sqrt n}$ for inverting a permutation.  The method
starts with choosing a set of pairs of inputs on which $f$ takes
different values.  Then the lower bound is determined by some
combinatorial properties of the graph of all pairs chosen.

Some functions, such as sorting or ordered search, could not be
satisfactorily lower-bounded by the unweighted adversary method.
H{\o}yer, Neerbek, and Shi used a novel
argument~\cite{hns:ordered-search} to obtain tight bounds for these
problems.  They weighted the input pairs and obtained the lower bound
by evaluating the spectral norm of the Hilbert matrix.
Barnum, Saks, and Szegedy proposed a general method~\cite{bss:semidef} that
gives necessary and sufficient conditions for the existence of a quantum query
algorithm.  They also described a special case, the so-called \emph{spectral
method}, which gives a lower bound in terms of spectral norms of an adversary
matrix.
Ambainis also published a \emph{weighted}
version of his adversary method~\cite{ambainis:degree-vs-qc}.  He
showed that it is stronger than the unweighted method and successfully
applied it to get a lower bound for several iterated functions.  This
method is slightly harder to apply, because it requires one to design
a so-called \emph{weight scheme}, which can be seen as a quantum
counterpart of the classical \emph{hard distribution} on the inputs.
Zhang observed that Ambainis had generalized his oldest
method~\cite{ambainis:lowerb} in two independent ways, so he unified them, and
published a \emph{strong weighted adversary method}~\cite{zhang:ambainis}.
Finally, Laplante and Magniez used Kolmogorov complexity in an unusual way and
described a \emph{Kolmogorov complexity method}~\cite{lm:kolmogorov-lb}.

All adversary lower-bound methods above except the Kolmogorov complexity
method were defined and proved only for Boolean functions, that is for
functions with Boolean input bits and a Boolean output.

A few relations between the methods are known.  It is a trivial fact that the
strong weighted adversary is at least as good as the weighted adversary.
Laplante and Magniez showed~\cite{lm:kolmogorov-lb} that the Kolmogorov
complexity method is at least as strong as all the following methods: the
Ambainis unweighted and weighted method, the strong weighted method, and the
spectral method.  The method of H\o yer et al.~\cite{hns:ordered-search} is a special case of
the weighted adversary method.  It seemed that there were several
incompatible variants of the quantum adversary method of different strength.

In addition it was known that there were some limitations for lower bounds obtained
by the adversary method.  Let $f$ be Boolean.
Szegedy observed~\cite{szegedy:triangle} that the
weighted adversary method is limited by $\min( \sqrt{C_0 n}, \sqrt{C_1
n})$, where $C_0$ is the zero-certificate complexity of $f$ and $C_1$ is the
one-certificate complexity of $f$.  Laplante and Magniez proved the same limitation for the Kolmogorov
complexity method~\cite{lm:kolmogorov-lb}, which implies that all other
methods are also bounded.
Finally, this bound was improved to $\sqrt{C_0 C_1}$ for total $f$ by
Zhang~\cite{zhang:ambainis} and independently by us.

\subsection{Our results}

In this paper, we clean up the forest of adversary methods.  First, we extend all
adversary lower bound methods to general non-Boolean functions.  Second, we show that
there is essentially only one quantum adversary method and that all the former
methods~\cite{%
bss:semidef,%
ambainis:degree-vs-qc,%
zhang:ambainis,%
lm:kolmogorov-lb}
are just different formulations of the
same method.  Since one method can be defined in several seemingly unrelated
ways and yet one always obtains the same bound, it implies that the quantum
adversary method is a very robust concept.

Third, we present a new simple proof of the limitation of the quantum
adversary method.  If we order the letters in the output alphabet by their
certificate complexities such that $C_0 \geq C_1 \geq \dots$, then all adversary
lower bounds are at most $2 \sqrt{C_1 n}$ for partial $f$ and $\sqrt{C_0 C_1}$
for total $f$.

\subsection{Separation between the polynomial and adversary method}

The polynomial method and the adversary method are generally incomparable.
There are examples when the polynomial method gives better bounds and vice
versa.

The polynomial method has been successfully applied to obtain tight lower
bounds for the following problems: $\Om{n^{1/3}}$ for the collision problem
and $\Om{n^{2/3}}$ for the element distinctness problem~\cite{as:collision}
(see~\cite{ambainis:eldist} for a matching upper bound).
The quantum adversary method is incapable of proving such lower bounds due to
the small certificate complexity of the function.  Furthermore, the polynomial
method often gives tight lower bounds for the exact and zero-error quantum
complexity, such as $n$ for the Or function~\cite{bbcmw:polynomialsj}.  The
adversary method completely fails in this setting and the only lower bound it
can offer is the bounded-error lower bound.

On the other hand, Ambainis exhibited some iterated
functions~\cite{ambainis:degree-vs-qc} for which the adversary method gives
better lower bounds than the polynomial method.  The
largest established gap between the two methods is $n^{1.321}$.
Furthermore, it is unknown how to apply the polynomial method to obtain
several lower bounds that are very simple to prove by the adversary
method. A famous example is the two-level And-Or tree.  The adversary
method gives a tight lower bound $\Om{\sqrt
  n}$~\cite{ambainis:lowerb}, whereas the best bound obtained by the
polynomial method is $\Om{n^{1/3}}$ and it
follows~\cite{ambainis:collision} from the element distinctness lower
bound~\cite{as:collision}.

There are functions for which none of the methods is known to give a tight
bound.  A long-standing open problem is the binary And-Or tree.  The best
known quantum algorithm is just an implementation of the classical zero-error
algorithm by Snir~\cite{snir:dec} running in expected time
$\O{n^{0.753}}$, which is optimal for both zero-error~\cite{sw:and-or} and
bounded-error~\cite{santha:and-or} algorithms.  The
adversary lower bounds are limited by $\sqrt{C_0 C_1} = \sqrt n$.
In a recent development, Laplante, Lee, and Szegedy
showed~\cite{lls:formulas} that this limitation $\sqrt n$ holds for every read-once
$\{ \land, \lor \}$ formula.  The best known
lower bound obtained by the polynomial method is also $\Om{\sqrt n}$ and it
follows from embedding the parity function.  It could be that the polynomial
method can prove a stronger lower bound.  Two other examples are triangle
finding and verification of matrix products.  For triangle finding, the best
upper bound is $\O{n^{1.3}}$~\cite{mss:triangle} and the best lower bound is
$\Om{n}$.  For verification of matrix products, the best upper bound is
$\O{n^{5/3}}$~\cite{bs:matrix} and the best lower bound is $\Om{n^{3/2}}$.
Again, the adversary method cannot give better bounds, but the polynomial
method might.


The semidefinite programming method~\cite{bss:semidef} gives an exact
characterization of quantum query complexity.  However, it is too
general to be applied directly.  It is an interesting open problem to
find a lower bound that cannot be proved by the adversary or polynomial
method.

\section{Preliminaries}

\subsection{Quantum query algorithms}

We assume familiarity with quantum computing~\cite{nielsen&chuang:qc}
and sketch the model of quantum query complexity,
referring to~\cite{buhrman&wolf:dectreesurvey} for more
details, also on the relation between query complexity
and certificate complexity.
Suppose we want to compute some function $f: S \to H$, where $S \subseteq G^N$
and $G, H$ are some finite alphabets.
For input $x\in S$, a \emph{query} gives us access to the
input variables. It corresponds to the unitary transformation, which depends on
input $x$ in the following way:
\[
O_x:\ket{i,b,z}\mapsto\ket{i,(b + x_i) \mod |G|,z} \enspace.
\]
Here $i\in[N]=\{1,\ldots,N\}$ and $b\in G$; the $z$-part corresponds
to the workspace, which is not affected by the query.
We assume the input can be accessed only via such queries.
A $T$-query quantum algorithm has the form $A=U_T O_x U_{T-1}\cdots O_x U_1
O_x U_0$,
where the $U_k$ are fixed unitary transformations, independent of $x$.
This $A$ depends on $x$ via the $T$ applications of $O_x$.
The algorithm starts in initial $S$-qubit state $\ket{0}$.
The output of $A$ is obtained by observing the first few qubits of the
final superposition $A\ket{0}$, and its
success probability on input $x$ is the probability of outputting $f(x)$.

\subsection{Kolmogorov complexity}

An excellent book about Kolmogorov complexity is the
book~\cite{li&vitanyi:kolm} by Li and Vit\'anyi.  Deep knowledge of Kolmogorov
complexity is not necessary to understand this paper.  Some results on the
relation between various classical forms of the quantum adversary method and
the Kolmogorov complexity method are taken from Laplante and
Magniez~\cite{lm:kolmogorov-lb}, and the others just use basic techniques.

A set is called \emph{prefix-free} if none of its members is a prefix of
another member.  Fix a universal Turing machine $M$ and a prefix-free set $S$.
The \emph{prefix-free Kolmogorov complexity} of $x$ given $y$, denoted by
$K(x|y)$, is the length of the shortest program from $S$ that prints $x$ if it
gets $y$ on the input.  Formally, 
\[
K(x|y) = \min \{ |P|: P \in S, M(P,y) = x\}\enspace.
\]

\subsection{Semidefinite programming}

In this paper, we use the duality theory of semidefinite programming
\cite{lovasz:semidef}.  There are various forms of
the duality principle in the literature.  We use a semidefinite extension of
Farkas's lemma~\cite[Theorem 3.4]{lovasz:semidef}.

\subsection{Notation}

Let $[n] = \{ 1, 2, \dots, n \}$.  Let $\Sigma^*$ denote the set of all finite
strings over alphabet $\Sigma$.  All logarithms are binary.
Let $I$ denote the \emph{identity} matrix.
Let $A^T$ denote the \emph{transpose} of $A$.
Let $\diag(A)$ denote the column vector containing the \emph{main diagonal} of $A$.
Let $\Tr(A)$ be the \emph{trace} of $A$
and let $A \cdot B$ be the scalar product of $A$ and $B$, formally $A \cdot B = \sum_{x,y} A[x,y] B[x,y]$.
For a column vector $x$, let $|x|$ denote the $\ell_2$-norm of $x$, formally $|x| = \sqrt{x^T x}$.
Let $\lambda(A)$ denote the \emph{spectral norm} of $A$, formally $\lambda(A) = \max_{x: |x| \neq 0} {|Ax| / |x|}$.
Let $A B$ denote the usual \emph{matrix product}
and let $A \circ B$ denote the \emph{Hadamard (point-wise) product}
\cite{mathias:spectral-norm}.
Formally, $(A B)[x,y] = \sum_i A[x,i] B[i,y]$
and $(A \circ B)[x,y] = A[x,y] B[x,y]$.
Let $A \geq B$ denote the \emph{point-wise comparison}
and let $C \succeq D$ denote that $C - D$ is \emph{positive semidefinite}.
Formally, $\forall x, y: A[x,y] \geq B[x,y]$
and $\forall v: v^T (C - D) v \geq 0$.
Let $\rownorm_x(M)$ denote the \emph{$\ell_2$-norm} of the $x$-th row of $M$
and let $\colnorm_y(M)$ denote the \emph{$\ell_2$-norm} of the $y$-th column
of $M$.
Formally, 
\[
\rownorm_x(M) = \sqrt{\sum_y M[x,y]^2} \qquad \text{and} \qquad \colnorm_y(M) = \sqrt{\sum_{\vphantom{y}x} M[x,y]^2}\enspace.
\]
Let $\rownorm(M) = \max_x \rownorm_x(M)$ and $\colnorm(M) = \max_y
\colnorm_y(M)$.

Let $S \subseteq G^n$ be a set of inputs.
We say that a function $f: S \to H$ is \emph{total}
if $S = G^n$.  A general function is called \emph{partial}.  Let $f$
be a partial function.  A \emph{certificate} for an input $x \in S$ is a
subset $I \subseteq [n]$ such that fixing the input variables $i \in I$ to $x_i$
determines the function value.  Formally, 
\[
\forall y \in S: y|_I = x|_I \Rightarrow f(y)
= f(x)\enspace,
\]
where $x|_I$ denotes the substring of $x$ indexed by $I$.  A
certificate $I$ for $x$ is called \emph{minimal}
if $|I| \leq |J|$ for every certificate $J$ for $x$.  Let $\Cert{x}$
denote the lexicographically smallest \emph{minimal certificate} for
$x$.  For an $h \in H$, let $C_h(f) = \max_{x: f(x)=h} |\Cert{x}|$ be
the \emph{$h$-certificate complexity} of $f$.

\section{Main result}

In this section, we present several equivalent quantum adversary methods and a
new simple proof of the limitations of these methods.  We can categorize these
methods into two groups.  Some of them solve conditions on the primal of the
quantum system~\cite{bss:semidef}: these are the spectral, weighted,
strong weighted, and generalized spectral adversary; and some of them solve
conditions on the dual: these are the Kolmogorov complexity bound, minimax,
and the semidefinite version of minimax.  Primal methods are mostly used to
give lower bounds on the query complexity, while we can use the duals to give
limitations of the method.

The primal methods, that is the spectral, weighted, and strong weighted
adversary, have been stated only for Boolean functions.  The generalization to
the more general non-Boolean case is straightforward and hence we state them
here in the generalized form.

\begin{theorem}
\label{thm:main}
Let $S \subseteq G^n$ and let $f: S \to H$ be a partial function.  Let
$Q_\eps(f)$ denote the $\eps$-error quantum query complexity of $f$.  Then
\[
\textstyle \frac {Q_\eps(f)} {1 - 2 \sqrt{\eps (1-\eps)}}
  \geq \SA = \WA = \SWA = \MM = \SMM = \GSA = \Th{\KA}\enspace,
\]
where {\rm SA, WA, SWA, MM, SMM, GSA,} and {\rm KA} are lower bounds given by
the following methods.
\begin{itemize}
\item {\bf Spectral adversary~\cite{bss:semidef}.}
Let $D_i, F$ be $|S| \times |S|$ zero-one valued matrices that satisfy $D_i[x,y] = 1$
iff $x_i \neq y_i$ for $i \in [n]$, and $F[x,y] = 1$ iff $f(x) \neq f(y)$.
Let $\Gamma$ denote an $|S| \times |S|$ non-negative symmetric matrix such that
$\Gamma \circ F = \Gamma$.
Then
\begin{equation}
\label{eqn:sa}
\SA = \max_\Gamma \frac {\lambda(\Gamma)} {\max_i \lambda(\Gamma \circ D_i)}
\enspace.
\end{equation}

\item {\bf Weighted adversary~\cite{ambainis:degree-vs-qc}.}%
\footnote{%
We use a different formulation~\cite{lm:kolmogorov-lb} than in the original
Ambainis papers~\cite{ambainis:lowerb,ambainis:degree-vs-qc}.  In particular,
we omit the relation $R
\subseteq A \times B$ on which the weights are required to be nonzero, and
instead allow zero weights.  It is simple to prove that both formulations are
equivalent.
}
Let $w, w'$ denote a weight scheme as follows:
\begin{itemize}
\item Every pair $(x,y) \in S^2$ is assigned a non-negative weight $w(x,y) =
w(y,x)$ that satisfies $w(x,y) = 0$ whenever $f(x) = f(y)$.
\item Every triple $(x,y,i) \in S^2 \times [n]$ is assigned a non-negative
weight $w'(x,y,i)$ that satisfies $w'(x,y,i) = 0$ whenever $x_i = y_i$ or
$f(x) = f(y)$, and $w'(x,y,i) w'(y,x,i) \geq w^2(x,y)$ for all $x,y,i$ such
that $x_i \neq y_i$ and $f(x) \neq f(y)$.  
\end{itemize}
For all $x,i$, let $wt(x) = \sum_y w(x,y)$ and $v(x,i) = \sum_y w'(x,y,i)$.
Then
\begin{equation}
\label{eqn:wa}
\WA = \max_{w,w'} \min_{\myatop {x,y,\ i,j} {\myatop {f(x) \neq f(y)} {v(x,i) v(y,j) > 0}}}
  \sqrt{\frac {wt(x) wt(y)} {v(x,i) v(y,j)}} \enspace.
\end{equation}

\item {\bf Strong weighted adversary~\cite{zhang:ambainis}.}
Let $w, w'$ denote a weight scheme as above.  Then
\begin{equation}
\label{eqn:swa}
\SWA = \max_{w,w'} \min_{\myatop {x,y,i} {\myatop {w(x,y) > 0} {x_i \neq y_i}}}
  \sqrt{\frac {wt(x) wt(y)} {v(x,i) v(y,i)}} \enspace.
\end{equation}

\item {\bf Kolmogorov complexity~\cite{lm:kolmogorov-lb}.}%
\footnote{%
We use a different formulation than Laplante and Magniez~\cite{lm:kolmogorov-lb}.  They
minimize over all algorithms $A$ computing $f$ and substitute $\sigma =
\mbox{source code of $A$}$, whereas we minimize over all finite strings
$\sigma$.  Our way is equivalent.  One can easily argue
that any finite string $\sigma$ can be ``embedded'' into any algorithm $B$: Let
$C$ be the source code of $B$ with appended comment $\sigma$ that is never
executed.  Now, the programs $B$ and $C$ are equivalent, and $K(x|\sigma) \leq
K(x|C) + \O1$ for every $x$.
}
Let $\sigma \in \01^*$ denote a finite string.  Then
\begin{equation}
\label{eqn:ka}
\KA = \min_\sigma \max_{\myatop {x,y} {f(x) \neq f(y)}}
  \frac 1 {\sum_{i: x_i \neq y_i} \sqrt{2^{-K(i|x,\sigma) -K(i|y,\sigma)}}}
  \enspace.
\end{equation}

\item {\bf Minimax over probability distributions~\cite{lm:kolmogorov-lb}.}
Let $p: S \times [n] \to \mathbb{R}$ denote a set of probability distributions, that is
$p_x(i) \geq 0$ and $\sum_i p_x(i) = 1$ for every $x$.  Then
\begin{eqnarray}
\MM =& &\min_p \max_{\myatop{x,y} {f(x) \neq f(y)}}
  \frac 1 {\sum_{i: x_i \neq y_i} \sqrt{ p_x(i)\, p_y(i) }}
  \label{eqn:minimax} \\
=& 1 \bigg/ &\max_p \min_{\myatop {x,y} {f(x) \neq f(y)}}
  \sum_{i: x_i \neq y_i} \sqrt{ p_x(i)\, p_y(i) } \enspace.
  \label{eqn:max-min}
\end{eqnarray}

\item {\bf Semidefinite version of minimax.}
Let $D_i, F$ be matrices as above.  Then
$\SMM = 1/\mu_{\rm max}$, where $\mu_{\rm max}$ is the maximal solution of the
following semidefinite program:
\begin{equation}
\label{eqn:smm}
\begin{array}{l @{\ } r @{\ } l}
\mbox{\rm maximize} & \omit $\mu$ \\
\mbox{\rm subject to}
  & \forall i: \hfil R_i & \succeq 0 \\
  & \sum_i R_i \circ I &= I \\
  & \sum_i R_i \circ D_i &\geq \mu F \enspace.
\end{array}
\end{equation}

\item {\bf Generalized spectral adversary.}
Let $D_i, F$ be matrices as above.  Then $\GSA = 1/\mu_{\rm min}$,
where $\mu_{\rm min}$ is the minimal solution of the following semidefinite
program:
\begin{equation}
\label{eqn:gsa}
\begin{array}{l @{\ } r @{\ } l}
\mbox{\rm minimize} & \omit $\mu = \Tr \Delta$\hfil \\
\mbox{\rm subject to}
  & \Delta & \rm is\ diagonal \\
  & Z &\geq 0 \\
  & Z \cdot F &= 1 \\
  & \forall i: \ \Delta - Z \circ D_i &\succeq 0 \enspace.
\end{array}
\end{equation}
\end{itemize}
\end{theorem}

Before we prove the main theorem in the next sections, let us draw some
consequences.  We show that there are limits that none of these
quantum adversary methods can go beyond.

\begin{theorem}
\label{thm:mini-max-bound}
Let $S \subseteq G^n$ and let $f: S \to H$ be a partial function.  Let the
output alphabet be $H = \{ 0, 1, \dots, |H|-1 \}$ and order the letters $h \in H$
by their $h$-certificate complexities such that $C_0 \geq C_1 \geq \dots \geq
C_{|H|-1}$.  Then the max-min bound~\eqnref{eqn:max-min} is upper-bounded by
$\MM \leq 2 \sqrt{C_1(f) \cdot n}$.  If $f$ is total, that is if $S = G^n$,
then $\MM \leq \sqrt{ C_0(f) \cdot C_1(f) }$.
\end{theorem}

\begin{proof}
The following simple argument is due to Ronald de Wolf.
We exhibit two sets of probability distributions $p$ such that
\[
m(p) = \min_{\myatop {x,y} {f(x) \neq f(y)}}
  \sum_{i: x_i \neq y_i} \sqrt{ p_x(i)\, p_y(i) }
  \geq \frac 1 {2 \sqrt{ C_1 n }} \enspace,
  \mbox{ resp. } \frac 1 {\sqrt{ C_0 C_1 }}\enspace.
\]
The max-min bound \eqnref{eqn:max-min} is $\MM = 1/ \max_p m(p)$ and the
statement follows.

Let $f$ be partial.  For every $x \in S$, distribute one half of the
probability uniformly over any minimal certificate $\Cert{x}$, and one half of
the probability uniformly over all input variables.  Formally, 
\[p_x(i) = \frac1{2n} + \frac1{2|\Cert{x}|}\; \text{iff}\; i \in
\Cert{x}\enspace,\qquad\text{and}\qquad p_x(i) = \frac1{2n}\; \text{for}\; i \not\in
\Cert{x}\enspace.
\]
Take any $x, y$ such that $f(x) \neq f(y)$.  Assume that $C_x \leq C_y$, and
take the $f(x)$-certificate $I = \Cert{x}$.  Since $y|_I \neq x|_I$,
there is a $j \in I$ such that $x_j \neq y_j$.  Now we lower-bound the sum
of~\eqnref{eqn:max-min}.
\[
\sum_{i: x_i \neq y_i} \sqrt{ p_x(i)\, p_y(i) }
  \geq \sqrt{ p_x(j)\, p_y(j) }
  \geq \sqrt{ \frac 1 {|2\Cert{x}|} \cdot \frac 1 {2n}}
  \geq \frac 1 {2 \sqrt{ C_{f(x)} n }}
  \geq \frac 1 {2 \sqrt{ C_1 n }}\enspace.
\]

Since this inequality holds for any $x, y$ such that $f(x) \neq f(y)$, also
$m(p) \geq 1 / 2 \sqrt{C_1 n}$.  Take the reciprocal and conclude that
$\MM \leq 2 \sqrt{C_1 n}$.

For Boolean output alphabet $H = \01$, we can prove a slightly stronger bound
$\MM \leq \sqrt{C_1 n}$ as follows.  Define $p$ as a uniform distribution over
some minimal certificate for all one-inputs, and a uniform distribution over
all input bits for all zero-inputs.  The same computation as above gives the
bound.

If $f$ is total, then we can do even better.  For every $x \in G^n$,
distribute the probability uniformly over any minimal certificate $\Cert{x}$.
Formally, $p_x(i) = 1/|\Cert{x}|$ iff $i \in \Cert{x}$, and $p_x(i) = 0$
otherwise.  Take any $x, y$ such that $f(x) \neq f(y)$, and let $I = \Cert{x}
\cap \Cert{y}$.  There must exist a $j \in I$ such that $x_j \neq y_j$, otherwise
we could find an input $z$ that is consistent with both certificates.  (That
would be a contradiction, because $f$ is total and hence $f(z)$ has to be
defined and be equal to both $f(x)$ and $f(y)$.)  After we have found a $j$, we
lower-bound the sum of~\eqnref{eqn:max-min} by $1 / \sqrt{C_{f(x)} C_{f(y)}}$ in
the same way as above.  Since $\sqrt{C_{f(x)} C_{f(y)}} \leq \sqrt{C_0 C_1}$,
the bound follows.
\end{proof}

Some parts of the following statement have been observed for
individual methods by
Szegedy \cite{szegedy:triangle}, Laplante and
Magniez~\cite{lm:kolmogorov-lb}, and Zhang~\cite{zhang:ambainis}.
This corollary rules out all adversary attempts to prove good lower bounds for
problems with small certificate complexity, such as element
distinctness~\cite{as:collision}, binary And-Or trees~\cite{%
bs:q-read-once,hmw:berror-search}, triangle finding~\cite{mss:triangle}, or
verification of matrix products~\cite{bs:matrix}.

\begin{corollary}
All quantum adversary lower bounds are at most
$\min( \sqrt{C_0(f) n},\penalty0 \sqrt{C_1(f) n})$ for partial Boolean functions and
$\sqrt{C_0(f) C_1(f)}$ for total Boolean functions.
\end{corollary}

\section{Equivalence of spectral and strong weighted adversary}

In this section, we give a linear-algebraic proof that the spectral
bound~\cite{bss:semidef} and the strong weighted bound~\cite{zhang:ambainis}
are equal.  The proof has three steps.  First, we show that the weighted
bound~\cite{ambainis:degree-vs-qc} is at least as good as the spectral bound.
Second, using a small combinatorial lemma, we show that the spectral bound is
at least as good as the strong weighted bound.  The strong weighted bound is
always at least as good as the weighted bound, because every term in the
minimization of \eqnref{eqn:swa} is included in the minimization of \eqnref{eqn:wa}:
if $w(x,y) > 0$ and $x_i \neq y_i$, then $f(x) \neq f(y)$ and both $w'(x,y,i) > 0$
and $w'(y,x,i) > 0$.  The generalization of the weighted adversary method thus
does not make the bound stronger, however its formulation is easier to use.

\subsection{Reducing spectral adversary to weighted adversary}

First, let us state two useful statements upper-bounding the spectral norm of
a Hadamard product of two non-negative matrices.  The first one is due to
Mathias~\cite{mathias:spectral-norm}.  The second one is our generalization
and its proof is postponed to \appref{app:spectral-norm}.

\begin{lemma}
\label{lem:spectral-norm}
{\rm \cite{mathias:spectral-norm}}
Let $S$ be a non-negative symmetric matrix and let $M$ and $N$ be
non-negative matrices such that $S \leq M \circ N$.  Then
\begin{equation}
\label{eqn:spectral-norm}
\lambda(S) \leq \rownorm(M) \colnorm(N)
  = \max_{x,y} \rownorm_x(M) \colnorm_y(N) \enspace.
\end{equation}
Moreover, for every symmetric $S \geq 0$ there exists an $M \geq 0$ such that $S
= M \circ M^T$ and $\rownorm(M) = \colnorm(M^T) = \sqrt{\lambda(S)}$.  This
optimal matrix can be written as $M[x,y] = \sqrt{S[x,y] \cdot d[y] / d[x]}$,
where $d$ is the principal eigenvector of $S$.
\end{lemma}

\begin{lemma}
\label{lem:cond-spectral-norm}
Let $S$ be a non-negative symmetric matrix and let $M$ and $N$ be
non-negative matrices such that $S \leq M \circ N$.  Then
\begin{equation}
\label{eqn:cond-spectral-norm}
\lambda(S) \leq \max_{\myatop {x,y} {S[x,y] > 0}} \rownorm_x(M) \colnorm_y(N)
\enspace.
\end{equation}
\end{lemma}

Now we use the first bound to reduce the spectral adversary to the weighted
adversary.

\begin{theorem}
\label{thm:spectral2weighted}
$\SA \leq \WA$.
\end{theorem}

\begin{proof}
Let $\Gamma$ be a non-negative symmetric matrix with $\Gamma \circ F =
\Gamma$ as in equation~\eqnref{eqn:sa} that gives the optimal spectral bound.
Assume without loss of generality that $\lambda(\Gamma) = 1$.  Let
$\delta$ be the principal eigenvector of $\Gamma$, that is $\Gamma \delta =
\delta$.  Define the following weight scheme:
\[
w(x,y) = w(y,x) = \Gamma[x,y] \cdot \delta[x] \delta[y]\enspace.
\]

Furthermore, for every $i$, using \lemref{lem:spectral-norm}, decompose
$\Gamma_i = \Gamma \circ D_i$ into a Hadamard product of two non-negative
matrices $\Gamma_i = M_i \circ M^T_i$ such that $\rownorm(M_i) =
\sqrt{\lambda(\Gamma_i)}$.  Define $w'$ as follows:
\[
w'(x,y,i) = M_i[x,y]^2 \delta[x]^2\enspace.
\]

Let us verify that $w, w'$ is a weight scheme.  From the definition, $w(x,y) =
w'(x,y,i) = 0$ if $f(x) = f(y)$, and also $w'(x,y,i) = 0$ if $x_i = y_i$.
Furthermore, if $f(x) \neq f(y)$ and $x_i \neq y_i$, then 
\[
w'(x,y,i) w'(y,x,i) = (M_i[x,y] \,\delta[x])^2 (M_i[y,x] \,\delta[y])^2 =
(\Gamma_i[x,y] \,\delta[x] \delta[y])^2 = w(x,y)^2\enspace.
\]

Finally, let us compute the bound~\eqnref{eqn:wa} given by the weight
scheme.
\begin{align*}
wt(x) &
  = \sum_y w(x,y)
  = \delta[x] \sum_y \Gamma[x,y] \delta[y]
  = \delta[x] \,(\Gamma \delta)[x]
  = \delta[x]^2 \enspace, \\
\frac {v(x,i)} {wt(x)} &
  = \frac {\sum_y w'(x,y,i)} {wt(x)}
  = \frac {\sum_y M_i[x,y]^2 \delta[x]^2} {\delta[x]^2}
  = \rownorm_x(M_i)^2
  \leq \rownorm(M_i)^2
  = \lambda(\Gamma_i) \enspace.
\end{align*}

The weighted adversary lower bound~\eqnref{eqn:wa} is thus at least
\[
\WA \geq \min_{\myatop {x,y,\ i,j} {\myatop {f(x) \neq f(y)} {v(x,i) v(y,j) > 0}}}
  \sqrt{\frac {wt(x) wt(y)} {v(x,i) v(y,j)}}
\geq \min_{i,j} \frac 1 {\sqrt{\lambda(\Gamma_i) \cdot \lambda(\Gamma_j)}}
= \frac {\lambda(\Gamma)} {\max_i \lambda(\Gamma_i)}
= \SA\enspace.
\]

Hence the weighted adversary is at least as strong as the spectral adversary
\eqnref{eqn:sa}.
\end{proof}

\subsection{Reducing strong weighted adversary to spectral adversary}

\begin{theorem}
\label{thm:strong-weighted2spectral}
$\SWA \leq \SA$.
\end{theorem}

\begin{proof}
Let $w, w'$ be a weight scheme as in Equation~\eqnref{eqn:wa} that gives the
optimal weighted bound.  Define the
following symmetric matrix $\Gamma$ on $S \times S$:
\[
\Gamma[x,y] = \frac {w(x,y)} {\sqrt{wt(x) wt(y)}}\enspace.
\]

We also define column vector $\delta$ on $S$ such that $\delta[x] =
\sqrt{wt(x)}$.  Let $W = \sum_x wt(x)$.  Then
\[
\lambda(\Gamma)
  \geq \delta^T \Gamma \delta / |\delta|^{2}
  = W/W = 1\enspace.
\]

Define the following matrix on the index set $S \times S$:
\[
M_i[x,y] = \sqrt{ \frac {w'(x,y,i)} {wt(x)} }\enspace.
\]

Every weight scheme satisfies $w'(x,y,i) w'(y,x,i) \geq w^{2}(x,y)$ for all
$x, y, i$ such that $x_i \neq y_i$.  Hence
\[
M_i[x,y] \cdot M_i[y,x]
  = \frac {\sqrt{w'(x,y,i) w'(y,x,i)}} {\sqrt{wt(x) wt(y)}}
  \geq \frac {w(x,y) \cdot D_i[x,y]} {\sqrt{wt(x) wt(y)}}
  = \Gamma_i[x,y]\enspace.
\]

This means that $\Gamma \leq M \circ M^T$.
By \lemref{lem:cond-spectral-norm} and using $c_y(M^T) = r_y(M)$,
\[
\lambda(\Gamma_i)
  \leq \max_{\myatop {x,y} {\Gamma_i[x,y]>0}} r_x(M) r_y(M)
= \max_{\myatop {x,y} {\myatop {w(x,y) > 0} {x_i \neq y_i}}} \sqrt{
  \sum_k \frac {w'(x,k,i)} {wt(x)}
  \sum_\ell \frac {w'(y,\ell,i)} {wt(y)}
  }
= \max_{\myatop {x,y} {\myatop {w(x,y) > 0} {x_i \neq y_i}}}
  \sqrt{\frac {v(x,i) v(y,i)} {wt(x) wt(y)}}\enspace.
\]

The spectral adversary lower bound~\eqnref{eqn:sa} is thus at least
\[
\SA
\geq \frac {\lambda(\Gamma)} {\max_i \lambda(\Gamma_i)}
\geq \min_i \min_{\myatop {x,y} {\myatop {w(x,y) > 0} {x_i \neq y_i}}}
  \sqrt{\frac {wt(x) wt(y)} {v(x,i) v(y,i)}}
= \SWA\enspace.
\]

Hence the spectral adversary is at least as strong as the weighted
adversary~\eqnref{eqn:swa}.
\end{proof}

\begin{remark}
The strength of the obtained reduction depends on which statement is used for
upper-bounding the spectral norm of $\Gamma_i$.
\begin{shortitemize}
\item
\lemref{lem:cond-spectral-norm} has just given us $\SWA \leq \SA$.
\item
\lemref{lem:spectral-norm} would give a weaker bound $\WA \leq \SA$.
\item
H{\o}yer, Neerbek, and Shi used an explicit expression for the norm of the
Hilbert matrix to get a lower bound for ordered
search~\cite{hns:ordered-search}.  Their method is thus also a special case of
the spectral method.
\item
Both versions of the original unweighted adversary
method~\cite{ambainis:lowerb} are obtained by using a spectral matrix $\Gamma$
corresponding to a zero-one valued weight scheme $w$, the
lower bound $\lambda(\Gamma) \geq d^T \Gamma d / |d|^2$, and
\lemref{lem:spectral-norm}, resp. \lemref{lem:cond-spectral-norm}.
\end{shortitemize}
\end{remark}

\section{Equivalence of minimax and generalized spectral adversary}

In this section, we prove that the minimax bound is equal to the
generalized spectral bound.  We first
remove the reciprocal by
taking the max-min bound.  Second, we write this bound as a
semidefinite program.  An application of duality theory of
semidefinite programming finishes the proof.

\begin{theorem}
\label{thm:maxmin2semidef}
$\MM = \SMM$.
\end{theorem}

\begin{proof}
Let $p$ be a set of probability distributions as in
Equation~\eqnref{eqn:max-min}.
Define $R_i[x,y] = \sqrt{ p_x(i)\, p_y(i) }$.  Since $p_x$ is a
probability distribution, we get that $\sum_i R_i$ must have all ones on the
diagonal.  The condition $\min_{\myatop {x,y} {f(x) \neq f(y)}} \sum_{i: x_i \neq y_i}
R_i[x,y] \geq \mu$
may be rewritten
\[\forall x,y: f(x) \neq f(y)
\Longrightarrow \sum_{i: x_i \neq y_i} R_i[x,y] \geq \mu\enspace,
\]
which is to say
$\sum_i R_i \circ D_i \geq \mu F$.  Each matrix $R_i$ should be an outer product
of a non-negative vector with itself: $R_i = r_i r_i^T$ for a column
vector $r_i[x] = \sqrt{p_x(i)}$.  We have, however, replaced that
condition by $R_i \succeq 0$ to get semidefinite program~\eqnref{eqn:smm}.
Since $r_i r_i^T \succeq 0$, the program~\eqnref{eqn:smm} is a relaxation of
the condition of~\eqnref{eqn:max-min} and $\SMM \leq \MM$.

Let us show that every solution $R_i$ of the semidefinite program can be
changed to an at least as good rank-1 solution $R'_i$.  Take a Cholesky
decomposition $R_i = X_i X_i^T$.  Define a column-vector $q_i[x] = \sqrt{ \sum_j
X_i[x,j]^2}$ and a rank-1 matrix $R'_i = q_i q_i^T$.  It is not hard to show
that all $R'_i$ satisfy the same constraints as $R_i$.  First, $R'_i$ is
positive semidefinite.  Second, $R'_i[x,x] = \sum_j X_i[x,j]^2 = R_i[x,x]$,
hence $\sum_i R'_i \circ I = I$.  Third, by a Cauchy-Schwarz inequality,
\[
R_i[x,y]
  = \sum_j X_i[x,j] X_i[y,j]
  \leq \sqrt{ \sum_k X_i[x,k]^2 } \sqrt{ \sum_\ell X_i[y,\ell]^2 }
  = q_i[x] q_i[y]
  = R'_i[x,y]\enspace,
\]
hence $\sum_i R'_i \circ D_i \geq \sum_i R_i \circ D_i \geq \mu F$.  We conclude
that $\MM \leq \SMM$.
\end{proof}

The equivalence of the semidefinite version of minimax and the generalized
spectral adversary is proved using the duality theory of semidefinite
programming.  We use a semidefinite version of Farkas's lemma
\cite[Theorem 3.4]{lovasz:semidef}.

\begin{theorem}
\label{thm:semidef2gen-spectral}
$\SMM = \GSA$.
\end{theorem}

\begin{proof}
Let us compute the dual of a semidefinite program without converting it
to/from the standard form, but using Lagrange multipliers.
Take the objective function $\mu$ of the semidefinite version of
minimax~\eqnref{eqn:smm} and add negative penalty terms for violating the
constraints.
\begin{eqnarray*}
\mu &+&
  \sum_i Y_i \cdot R_i
  + D \cdot \Big( \sum_i R_i \circ I - I \Big)
  + Z \cdot \Big( \sum_i R_i \circ D_i - \mu F \Big) = \\
&& \mbox{for $Y_i \succeq 0$, unconstrained $D$, and $Z \geq 0$} \\
&=& \sum_i R_i \cdot \Big( Y_i + D \circ I + Z \circ D_i \Big)
  + \mu \Big( 1 - Z \cdot F \Big) - D \cdot I \enspace.
\end{eqnarray*}
Its dual system is formed by the constraints on $Y_i$, $D$, and $Z$ plus the
requirements that both expression in the parentheses are zero.  The duality
principle~\cite[Theorem 3.4]{lovasz:semidef} says that any primal solution is
smaller than or equal to any dual solution.  Moreover, if any of the two
systems has a strictly feasible solution, then the maximal primal solution
equals to the minimal dual solution.

Since $Y_i \succeq 0$ only appears once, we get rid of it by requiring that $D
\circ I + Z \circ D_i \preceq 0$.  We substitute $\Delta = -D \circ I$ and
obtain $\Delta - Z \circ D_i \succeq 0$.  The objective function is $-D \cdot
I = \Tr \Delta$.  We have obtained the generalized spectral
adversary~\eqnref{eqn:gsa}.  Let us prove its strong feasibility.  Assume that
the function $f$ is not constant, hence $F \neq 0$.  Take $Z$ a uniform
probability distribution over nonzero entries of $F$ and a large enough
constant $\Delta$.  This is a strictly feasible solution.  We conclude that
$\mu_{\rm max} = \mu_{\rm min}$.
\end{proof}

\section{Equivalence of generalized spectral and spectral adversary}

In this section, we prove that the generalized spectral adversary bound is
equal to the spectral adversary bound.  The main difference between
them is that the generalized method uses an arbitrary positive diagonal matrix
$\Delta$ as a new variable instead of the identity matrix $I$.

\begin{theorem}
\label{thm:gen-spectral2spectral}
$\GSA = \SA$.
\end{theorem}

\begin{proof}
Let $Z, \Delta$ be a solution of~\eqnref{eqn:gsa}.
First, let us prove that $\Delta \succ 0$.  Since both $Z \geq 0$ and $D_i
\geq 0$, it holds that $\diag(-Z \circ D_i) \leq 0$.  We know that $\Delta - Z
\circ D_i \succeq 0$, hence $\diag(\Delta - Z \circ D_i) \geq 0$, and
$\diag(\Delta) \geq 0$ follows.  Moreover, $\diag(\Delta) > 0$ unless $Z$
contains an empty row, in which case we delete it (together with the
corresponding column) and continue.  Second, since positive semidefinite
real matrices are symmetric, $\Delta - Z \circ D_i \succeq 0$
implies that $Z \circ D_i$ is symmetric (for every $i$).  For every $x \neq
y$ there is a bit $i$ such that $x_i \neq y_i$, hence $Z$
must be also symmetric.

Take a column vector $a = \diag(\Delta^{-1/2})$ and a rank-1
matrix $A = a a^T \succ 0$.  It is simple to prove that $A \circ X \succeq 0$
for every matrix $X \succeq 0$.
\rem{
\newcommand{\Diag}{\mathinner{\mathrm{Diag}}}
\[
v^T (A \circ X)\, v
  = v^T \Diag(a) X \Diag(a)\, v
  = w^T X w
  \geq 0\enspace.
\]
}
Since $\Delta - Z \circ D_i \succeq 0$,
also $A \circ (\Delta - Z \circ D_i) = I - Z \circ D_i \circ A \succeq 0$ and
hence $\lambda(Z \circ D_i \circ A) \leq 1$.
Now, define the spectral adversary matrix $\Gamma = Z \circ F \circ A$.
Since $0 \leq Z \circ F \leq Z$, it follows that
\[
\lambda(\Gamma \circ D_i) 
  = \lambda( Z \circ F \circ A \circ D_i)
  \leq \lambda( Z \circ D_i \circ A)
  \leq 1\enspace.
\]
It remains to show that $\lambda(\Gamma) \geq 1/\Tr \Delta$.  Let $b = \diag(\sqrt
\Delta)$ and $B = b b^T$.  Then
\[
1
  = Z \cdot F 
  = \Gamma \cdot B
  = b^T \Gamma b
  \leq \lambda(\Gamma) \cdot |b|^2
  = \lambda(\Gamma) \cdot \Tr \Delta\enspace.
\]
It is obvious that $\Gamma$ is symmetric, $\Gamma \geq 0$, and $\Gamma \circ F
= \Gamma$.  The bound~\eqnref{eqn:sa} given by $\Gamma$ is bigger than or equal
to $1/\Tr \Delta$, hence $\SA \geq \GSA$.

For the other direction, let $\Gamma$ be a non-negative symmetric matrix
satisfying $\Gamma \circ F = \Gamma$.  Let $\delta$ be its principal
eigenvector with $|\delta|=1$.  Assume without loss of generality that
$\lambda(\Gamma) = 1$ and let $\mu = \max_i \lambda(\Gamma_i)$.  Take $A =
\delta \delta^T$, $Z = \Gamma \circ A$, and $\Delta = \mu I \circ A$.  Then $Z
\cdot F = \Gamma \cdot A = \delta^T \Gamma \delta = 1$ and $\Tr \Delta = \mu$.
For every $i$, $\lambda(\Gamma_i) \leq \mu$, hence $\mu I - \Gamma \circ D_i
\succeq 0$.  It follows that $0 \preceq A \circ (\mu I - \Gamma \circ D_i) =
\Delta - Z \circ D_i$.  The semidefinite program~\eqnref{eqn:gsa} is satisfied
and hence its optimum is $\mu_{\rm min} \leq \mu$.  We conclude that $\GSA \geq
\SA$.
\end{proof}

\section{Proof of the main theorem}

In this section, we close the circle of reductions.  We use the results of
Laplante and Magniez, who recently proved~\cite{lm:kolmogorov-lb} that the
Kolmogorov complexity bound is asymptotically lower-bounded by the weighted
adversary bound and upper-bounded by the minimax bound.  The upper bound is
implicit in their paper, because they did not state the minimax bound as a
separate theorem.

\begin{theorem}
\label{thm:adversary2kolmogorov}
{\rm \cite[Theorem 2]{lm:kolmogorov-lb}}
$\KA = \Om{\WA}$.
\end{theorem}

\begin{theorem}
\label{thm:kolmogorov2minimax}
$\KA = \O{\MM}$.
\end{theorem}

\begin{proof}
Take a set of probability distributions $p$ as in Equation~\eqnref{eqn:minimax}.
The query information lemma \cite[Lemma 3]{lm:kolmogorov-lb} says that $K(i|x,p)
\leq \log \frac 1 {p_x(i)} +
\O1$ for every $x,i$ such that $p_x(i) > 0$.  This is true, because any $i$ of
nonzero probability can be encoded in $\lceil \log \frac 1 {p_x(i)} \rceil$ bits
using the Shannon-Fano code of distribution $p_x$, and the Shannon-Fano code
is a prefix-free code.  Rewrite the inequality as $p_x(i) = \O{2^{-K(i|x,p)}}$.
The statement follows, because the set of all strings $\sigma$
in~\eqnref{eqn:ka} includes among others also
the descriptions of all probability distributions $p$.
\end{proof}

\begin{remark}
The constant in the equality $\KA = \Th{\WA}$ depends on the choice of the
universal Turing machine and the prefix-free set.
\end{remark}

\begin{proof}[Proof of \thmref{thm:main}.]
We have to prove that
\[
\frac {Q_\eps(f)} {1 - 2 \sqrt{\eps (1-\eps)}}
  \geq \SA = \WA = \SWA = \MM = \SMM = \GSA = \Th{\KA}\enspace.
\]
Put together all known equalities and inequalities.
\begin{shortitemize}
\item $\SA = \WA = \SWA$ by \thmref{thm:spectral2weighted} and \thmref{thm:strong-weighted2spectral},
\item $\MM = \SMM$ by \thmref{thm:maxmin2semidef},
\item $\SMM = \GSA$ by \thmref{thm:semidef2gen-spectral},
\item $\GSA = \SA$ by \thmref{thm:gen-spectral2spectral},
\item $\KA = \Th{\WA}$ by \thmref{thm:adversary2kolmogorov} and \thmref{thm:kolmogorov2minimax}.
\end{shortitemize}
Finally, one has to prove one of the lower bounds.  For example, Ambainis
proved~\cite{ambainis:degree-vs-qc} that $Q_2(f) \geq (1 - 2 \sqrt{\eps
(1-\eps)}) \,\WA$ for every Boolean $f$.  Laplante and
Magniez proved~\cite{lm:kolmogorov-lb} that $Q_2(f) = \Om{\KA}$ for general
$f$.  H\o yer and \v Spalek present in their survey~\cite{hs:survey-lb} an
alternative proof of the spectral adversary bound that can easily be adapted
to the non-Boolean case.
\end{proof}

\appendix

\section{Proof of the upper bound on the spectral norm}
\label{app:spectral-norm}

\begin{proof}[Proof of \lemref{lem:cond-spectral-norm}.]
Let $S = M \circ N$.  Define a shortcut
\[
B(M,N) = \max_{\myatop {x,y} {S[x,y] > 0}} \rownorm_x(M) \colnorm_y(N)\enspace.
\]
Without loss of generality, we assume that $M[x,y] = 0 \Leftrightarrow N[x,y]
= 0 \Leftrightarrow S[x,y] = 0$.  Let us prove the existence of matrices $M',
N'$ with $B(M',N') = \rownorm(M') \colnorm(N')$ such that
\begin{equation}
\label{eqn:bmn}
M \circ N = M' \circ N', \mbox{ and } B(M, N) = B(M', N') \enspace.
\end{equation}
We then apply \lemref{lem:spectral-norm} and obtain
\[
\lambda(S)
  \leq \lambda(M \circ N)
  = \lambda(M' \circ N')
  \leq \rownorm(M') \colnorm(N')
  = B(M', N')
  = B(M, N)\enspace.
\]
{%
\newcommand{\CC}{{\overline C}}%
\newcommand{\RR}{{\overline R}}%
Take as $M', N'$ any pair of matrices that satisfies \eqnref{eqn:bmn} and the following
constraints:
\begin{itemize}
\item
$b = \rownorm(M') \colnorm(N')$ is minimal, that is there is no pair $M'',
N''$ giving a smaller $b$,
\item and, among those, the set $R$ of maximum-norm rows of $M'$ and the set
$C$ of maximum-norm columns of $N'$ are both minimal (in the same sense).
\end{itemize}
Let $(r,c)$ be any ``maximal'' entry, that is
$S[r,c] > 0$ and $\rownorm_r(M') \colnorm_c(N')
= B(M',N')$.  Let $\RR$ denote the \emph{complement} of $R$ and let
$S[R,C]$ denote the \emph{sub-matrix} of $S$ indexed by $R \times C$.  Then one
of the following cases happens:
\begin{enumerate}
\item $(r,c) \in R \times C$:
Then $B(M',N') = \rownorm(M') \colnorm(N')$ and we are done.
If this is not the case, then we know that $S[R,C] = 0$.

\item $(r,c) \in R \times \CC$:
Then $S[\RR,C] = 0$, otherwise we get a contradiction with one of the
minimality assumptions.  If $S[x,y] \neq 0$ for some $(x,y) \in \RR \times C$,
multiply $M'[x,y]$ by $1 + \eps$ and divide $N'[x,y]$ by $1 + \eps$ for some
small $\eps > 0$ such that the norm of the $x$-th row of $M'$ is still smaller
than $\rownorm(M')$.  Now, we have either deleted the $y$-th column from $C$ or,
if $|C|=1$, decreased $\colnorm(N')$.  Both cases are a contradiction.
Finally, if $S[\RR,C] = 0$, then $\colnorm(N') = 0$ due to $S[R,C] = 0$ and
the fact that $C$ are the maximum-norm columns.  Hence $S$ is a zero matrix,
and we are done.

\item $(r,c) \in \RR \times C$:
This case is similar to the previous case.

\item $(r,c) \in \RR \times \CC$:
First, note that $S[R,c] = 0$, otherwise $(r,c)$ would not be ``maximal''.
Now we divide all entries in $M'[R,\CC]$ by $1 + \eps$ and multiply all
entries in $N'[R,\CC]$ by $1 + \eps$ for some small $\eps > 0$ such that the
``maximal'' entries are unchanged.  Since $S[R,C]=0$, it follows that either
$S[R,\CC]=0$ and $S$ is a zero matrix, or there is a nonzero number in every
row of $M'[R,\CC]$.  Therefore, unless $S$ is a zero matrix, we have
preserved $B(M',N')$ and $\colnorm(N')$, and decreased $\rownorm(M')$, which
is a contradiction.
\end{enumerate}
We conclude that $(r,c) \in R \times C$, $B(M',N') = \rownorm(M')
\colnorm(N')$, and hence $\lambda(S) \leq B(M,N)$.
}
\end{proof}

\section*{Acknowledgments}

We thank Ronald de~Wolf for many fruitful discussions, for his suggestions
concerning \thmref{thm:mini-max-bound}, and for proofreading,
and Troy Lee for discussions.  We thank anonymous referees for their helpful
comments.

Robert {\v S}palek is supported in part by the EU fifth framework project
RESQ, IST-2001-37559.  Mario Szegedy is supported by NSF grant 0105692, and in
part by the National Security Agency (NSA) and Advanced Research and
Development Activity (ARDA) under Army Research Office (ARO), contract number
DAAD19-01-1-0506.

\bibliographystyle{tocplain}
\bibliography{equiv}

\providecommand{\bibhead}[1]{}
\expandafter\ifx\csname pdfbookmark\endcsname\relax%
  \providecommand{\tocrefpdfbookmark}{}
\else\providecommand{\tocrefpdfbookmark}{%
   \hypertarget{tocreferences}{}%
   \pdfbookmark[1]{References}{tocreferences}}%
\fi

\tocrefpdfbookmark
\begin{thebibliography}{10}

\bibitem{as:collision}\bibhead{as:collision}
{\sc S.~Aaronson and Y.~Shi}: Quantum lower bounds for the collision and the
  element distinctness problem.
\newblock {\em Journal of the ACM}, 51(4):595--605, 2004.
\newblock [\epfmt{jacm}{1008731.1008735}, \epfmt{arxiv}{quant-ph/0111102}].

\bibitem{ambainis:lowerb}\bibhead{ambainis:lowerb}
{\sc A.~Ambainis}: Quantum lower bounds by quantum arguments.
\newblock {\em Journal of Computer and System Sciences}, 64(4):750--767, 2002.
\newblock Earlier version in STOC '00.
\newblock [\epfmt{jcss}{10.1006/jcss.2002.1826}, \epfmt{stoc}{335305.335394},
  \epfmt{arxiv}{quant-ph/0002066}].

\bibitem{ambainis:degree-vs-qc}\bibhead{ambainis:degree-vs-qc}
{\sc A.~Ambainis}: Polynomial degree {vs.} quantum query complexity.
\newblock In {\em Proc. of 44th IEEE FOCS}, pp. 230--239, 2003.
\newblock [\epfmt{focs}{2003.1238197}, \epfmt{arxiv}{quant-ph/0305028}].

\bibitem{ambainis:eldist}\bibhead{ambainis:eldist}
{\sc A.~Ambainis}: Quantum walk algorithm for element distinctness.
\newblock In {\em Proc. of 45th IEEE FOCS}, pp. 22--31, 2004.
\newblock [\epfmt{focs}{2004.54}, \epfmt{arxiv}{quant-ph/0311001}].

\bibitem{ambainis:collision}\bibhead{ambainis:collision}
{\sc A.~Ambainis}: Polynomial degree and lower bounds in quantum complexity:
  Collision and element distinctness with small range.
\newblock {\em Theory of Computing}, 1:37--46, 2005.
\newblock [\epfmt{toc}{v001/a003}, \epfmt{arxiv}{quant-ph/0305179}].

\bibitem{bs:q-read-once}\bibhead{bs:q-read-once}
{\sc H.~Barnum and M.~Saks}: A lower bound on the quantum query complexity of
  read-once functions.
\newblock {\em Journal of Computer and System Sciences}, 69(2):244--258, 2004.
\newblock [\epfmt{jcss}{10.1016/j.jcss.2004.02.002},
  \epfmt{arxiv}{quant-ph/0201007}].

\bibitem{bss:semidef}\bibhead{bss:semidef}
{\sc H.~Barnum, M.~Saks, and M.~Szegedy}: Quantum decision trees and
  semidefinite programming.
\newblock In {\em Proc. of 18th IEEE Complexity}, pp. 179--193, 2003.
\newblock [\epfmt{ccc}{10.1109/CCC.2003.1214419}].

\bibitem{bbcmw:polynomialsj}\bibhead{bbcmw:polynomialsj}
{\sc R.~Beals, H.~Buhrman, R.~Cleve, M.~Mosca, and R.~{de} Wolf}: Quantum lower
  bounds by polynomials.
\newblock {\em Journal of the ACM}, 48(4):778--797, 2001.
\newblock Earlier version in FOCS '98.
\newblock [\epfmt{jacm}{502090.502097}, \epfmt{focs}{1998.743485},
  \epfmt{arxiv}{quant-ph/9802049}].

\bibitem{bbbv:hybrid}\bibhead{bbbv:hybrid}
{\sc H.~Bennett, E.~Bernstein, G.~Brassard, and U.~Vazirani}: Strengths and
  weaknesses of quantum computing.
\newblock {\em SIAM Journal on Computing}, 26(5):1510--1523, 1997.
\newblock [\epfmt{sicomp}{30093}, \epfmt{arxiv}{quant-ph/9701001}].

\bibitem{bs:matrix}\bibhead{bs:matrix}
{\sc H.~Buhrman and R.~{\v{S}}palek}: Quantum verification of matrix products.
\newblock In {\em Proc. of 17th ACM-SIAM SODA}, pp. 880--889, 2006.
\newblock [\epfmt{soda}{1109557.1109654}, \epfmt{arxiv}{quant-ph/0409035}].

\bibitem{buhrman&wolf:dectreesurvey}\bibhead{buhrman&wolf:dectreesurvey}
{\sc H.~Buhrman and R.~{de} Wolf}: Complexity measures and decision tree
  complexity: {A} survey.
\newblock {\em Theoretical Computer Science}, 288(1):21--43, 2002.
\newblock [\epfmt{tcs}{10.1016/S0304-3975(01)00144-X}].

\bibitem{grover:search}\bibhead{grover:search}
{\sc L.~K. Grover}: A fast quantum mechanical algorithm for database search.
\newblock In {\em Proc. of 28th ACM STOC}, pp. 212--219, 1996.
\newblock [\epfmt{stoc}{237814.237866}, \epfmt{arxiv}{quant-ph/9605043}].

\bibitem{hmw:berror-search}\bibhead{hmw:berror-search}
{\sc P.~H{\o}yer, M.~Mosca, and R.~{de} Wolf}: Quantum search on bounded-error
  inputs.
\newblock In {\em Proc. of 30th ICALP}, pp. 291--299, 2003.
\newblock LNCS 2719.
\newblock [\epfmt{icalp}{214dhep41d6vk3d2}, \epfmt{arxiv}{quant-ph/0304052}].

\bibitem{hns:ordered-search}\bibhead{hns:ordered-search}
{\sc P.~H{\o}yer, J.~Neerbek, and Y.~Shi}: Quantum complexities of ordered
  searching, sorting, and element distinctness.
\newblock {\em Algorithmica}, 34(4):429--448, 2002.
\newblock Special issue on Quantum Computation and Cryptography.
\newblock [\epfmt{algorithmica}{25gl9elr5rxr3q6a},
  \epfmt{arxiv}{quant-ph/0102078}].

\bibitem{hs:survey-lb}\bibhead{hs:survey-lb}
{\sc P.~H{\o}yer and R.~{\v{S}}palek}: Lower bounds on quantum query
  complexity.
\newblock {\em EATCS Bulletin}, 87:78--103, October, 2005.
\newblock [\epfmt{arxiv}{quant-ph/0509153}].

\bibitem{lls:formulas}\bibhead{lls:formulas}
{\sc S.~Laplante, T.~Lee, and M.~Szegedy}: The quantum adversary method and
  formula size lower bounds.
\newblock In {\em Proc. of 20th IEEE Complexity}, pp. 76--90, 2005.
\newblock [\epfmt{ccc}{10.1109/CCC.2004.1313852},
  \epfmt{arxiv}{quant-ph/0501057}].

\bibitem{lm:kolmogorov-lb}\bibhead{lm:kolmogorov-lb}
{\sc S.~Laplante and F.~Magniez}: Lower bounds for randomized and quantum query
  complexity using {Kolmogorov} arguments.
\newblock In {\em Proc. of 19th IEEE Complexity}, pp. 294--304, 2004.
\newblock [\epfmt{ccc}{10.1109/CCC.2004.1313852},
  \epfmt{arxiv}{quant-ph/0311189}].

\bibitem{li&vitanyi:kolm}\bibhead{li&vitanyi:kolm}
{\sc M.~Li and P.~M.~B. Vit{\'{a}}nyi}: {\em An Introduction to {K}olmogorov
  Complexity and its Applications}.
\newblock Springer, Berlin, second edition, 1997.

\bibitem{lovasz:semidef}\bibhead{lovasz:semidef}
{\sc L.~Lov\'asz}: Semidefinite programs and combinatorial optimization.
\newblock \url{http://research.microsoft.com/users/lovasz/semidef.ps}, 2000.

\bibitem{mss:triangle}\bibhead{mss:triangle}
{\sc F.~Magniez, M.~Santha, and M.~Szegedy}: Quantum algorithms for the
  triangle problem.
\newblock In {\em Proc. of 16th ACM-SIAM SODA}, pp. 1109--1117, 2005.
\newblock [\epfmt{soda}{1070432.1070591}, \epfmt{arxiv}{quant-ph/0310134}].

\bibitem{mathias:spectral-norm}\bibhead{mathias:spectral-norm}
{\sc R.~Mathias}: The spectral norm of a nonnegative matrix.
\newblock {\em Linear Algebra and its Applications}, 139:269--284, 1990.
\newblock [\epfmt{laa}{10.1016/0024-3795(90)90403-Y}].

\bibitem{nielsen&chuang:qc}\bibhead{nielsen&chuang:qc}
{\sc M.~A. Nielsen and I.~L. Chuang}: {\em Quantum Computation and Quantum
  Information}.
\newblock Cambridge University Press, 2000.

\bibitem{sw:and-or}\bibhead{sw:and-or}
{\sc M.~Saks and A.~Wigderson}: Probabilistic {Boolean} decision trees and the
  complexity of evaluating games trees.
\newblock In {\em Proc. of 27th IEEE FOCS}, pp. 29--38, 1986.

\bibitem{santha:and-or}\bibhead{santha:and-or}
{\sc M.~Santha}: On the {M}onte {C}arlo decision tree complexity of read-once
  formulae.
\newblock {\em Random Structures and Algorithms}, 6(1):75--87, 1995.

\bibitem{snir:dec}\bibhead{snir:dec}
{\sc M.~Snir}: Lower bounds on probabilistic decision trees.
\newblock {\em Theoretical Computer Science}, 38:69--82, 1985.
\newblock [\epfmt{tcs}{10.1016/0304-3975(85)90210-5}].

\bibitem{szegedy:triangle}\bibhead{szegedy:triangle}
{\sc M.~Szegedy}: On the quantum query complexity of detecting triangles in
  graphs.
\newblock quant-ph/0310107, 2003.
\newblock [\epfmt{arxiv}{quant-ph/0310107}].

\bibitem{zhang:ambainis}\bibhead{zhang:ambainis}
{\sc S.~Zhang}: On the power of {Ambainis}'s lower bounds.
\newblock {\em Theoretical Computer Science}, 339(2--3):241--256, 2005.
\newblock Earlier version in ICALP'04.
\newblock [\epfmt{tcs}{10.1016/j.tcs.2005.01.019},
  \epfmt{icalp}{gm2ff6wpc0q39v3x}, \epfmt{arxiv}{quant-ph/0311060}].

\end{thebibliography}

\pagebreak

\begin{tocauthors}
\begin{tocinfo}[spalek]%
  Robert {\v S}palek \\
  graduate student \\
  Centrum voor Wiskunde en Informatica\\
  Amsterdam, The Netherlands\\
  sr\tocat{}cwi\tocdot{}nl \\
  \url{http://www.ucw.cz/~robert/}
\end{tocinfo}
\begin{tocinfo}[szegedy]%
  Mario Szegedy \\
  professor \\
  Rutgers, the State University of New Jersey \\
  Piscataway, New Jersey, USA \\
  szegedy\tocat{}cs\tocdot{}rutgers\tocdot{}edu \\
  \url{http://athos.rutgers.edu/~szegedy/}
\end{tocinfo}
\end{tocauthors}

\begin{tocaboutauthors}
\begin{tocabout}[spalek] {\sc Robert \v Spalek} received his Masters
  Degrees in Computer Science from
  \href{http://www.mff.cuni.cz/}{Charles University}, Prague and
  \href{http://www.cs.vu.nl/}{Vrije Universiteit}, Amsterdam.  He is
  currently a graduate student at
  \href{http://www.cwi.nl/}{CWI}, advised by
  \href{http://homepages.cwi.nl/~buhrman/}{Harry Buhrman}.  His
  research interests include quantum computing, computational
  complexity, algorithms, data structures, and search engines.  He
  loves dancing salsa, climbing, photography, travelling to distant
  countries, and playing guitar.
\end{tocabout}

\begin{tocabout}[szegedy] {\sc Mario Szegedy} received his Ph.\,D. in
  computer science at the \href{http://www.cs.uchicago.edu}{University
    of Chicago} under the supervision of
  \href{http://www.cs.uchicago.edu/~laci}{Laci Babai} and
  \href{http://www.cs.uchicago.edu/~simon}{Janos Simon}.  He held
  a Lady Davis Postdoctoral Fellowship at the 
  \href{http://www.huji.ac.il/}{Hebrew University}, Jerusalem (1989-90),
  a postdoc at the \href{http://www.cs.uchicago.edu}{University of Chicago},
  1991-92, and a postdoc at \href{http://www.bell-labs.com/}{Bell
  Laboratories} (1992). He was a permanent member of Bell Labs for 7 years
  and for two more years of \href{http://public.research.att.com/}{AT\&T
    Research}. He left AT\&T in September 1999 to conduct research at
  the \href{http://www.ias.edu}{Institute for Advanced Study} in
  Princeton for a year.    In 2000 he joined the faculty of
  \href{http://www.rutgers.edu/}{Rutgers University}. 

  \noindent
  He received the \href{http://sigact.acm.org/prizes/godel/}{G\"{o}del
    Prize} twice, in
  \href{http://sigact.acm.org/prizes/godel/2001.html}{2001} for his
  part in the PCP Theorem and its connection to inapproximability and
  in \href{http://sigact.acm.org/prizes/godel/2005.html}{2005} for the
  analysis of data streams using limited memory.

  \noindent
  His research interests include complexity theory, combinatorics,
  combinatorial geometry and quantum computing, but he also has an
  interest in algebra and in programming languages.

  \noindent
  With a group of students he has founded QCteam, a quantum computing
  laboratory at Rutgers, which is his main project at the present
  time.  The laboratory has received substantial funding from the
  university and from the National Science Foundation. It has a
  vigorous visitor program, and pursues collaboration with the local
  industry.
\end{tocabout}
\end{tocaboutauthors}

\end{document}